\documentclass[aps, pra,11pt,showpacs]{revtex4-1}
\usepackage{amsmath}    
\usepackage{graphicx}   
\usepackage{verbatim}   
\usepackage{color}      
\usepackage{subfigure}  
\usepackage{hyperref}   
\usepackage{dcolumn}
\begin{document}

\title{Dynamics of laterally-coupled pairs of spin-VCSELs.}

\author{M.P. Vaughan$^{1}$}\email[Corresponding author: ]{mpvaug@essex.ac.uk}
\author{H. Susanto$^{2}$}
\author{I.D. Henning$^{1}$}
\author{M.J. Adams$^{1}$}

\affiliation{$^{1}$School of Computer Science and Electronic Engineering, University of Essex, Wivenhoe Park, Colchester CO4 3SQ, United Kingdom}

\affiliation{$^{2}$Department of Mathematical Sciences, University of Essex, Wivenhoe Park, Colchester CO4 3SQ, United Kingdom}

\date{\today}

\begin{abstract}
A newly-developed normal mode model of laser dynamics in a generalised array of waveguides is applied to extend the spin-flip model (SFM) to pairs of evanescently-coupled spin-VCSELS. The effect of high birefringence is explored, revealing new dynamics and regions of bistability. It is shown that optical switching of the polarisation states of the lasers may be controlled through the optical pump and that, under certain conditions, the polarisation of one laser may be switched by controlling the intensity and polarisation in the other.
\end{abstract}

\maketitle

\section{Introduction}
Recent years have seen a growth of research interest in the nonlinear dynamics of arrays of vertical cavity surface-emitting lasers (VCSELs) and in potential applications of these effects. Notable advances include work on parity-time symmetry and non-Hermiticity associated with the control of gain and loss in neighbouring VCSEL cavities \cite{ gao2017parity, gao2018rate, gao2019non,  dave2019static}. Progress has also been rapid in the understanding of ultrahigh-speed resonances that offer the prospect of very high frequency modulation of coupled VCSELs and nanolasers \cite{fryslie2017modulation, xiao2017modulation, han2018analysis, kominis2019antiresonances}. Additional insight into optical coupling between adjacent elements of a two-dimensional VCSEL array has been achieved by careful analysis of the effects of varying the injected current independently on each array element \cite{ pan2018analysis}. The coupling was shown to provide extra optical gain for array elements and thus lead to additional output power of the array due to in-phase operation \cite{pan2018analysis, pan2018large}, reduced thresholds of individual elements \cite{pan2018analysis, pan2019ultra} and even cause unpumped elements to lase \cite{pan2018analysis}.

In almost all the above examples of recent progress, modelling of the array behaviour based on coupled mode theory (CMT) has been used to explain experimental results and develop improved understanding of fundamental effects. Conventional CMT describes only the amplitude and phase of the electric field of the photons and the total concentration of the electrons. Whilst this is adequate for modelling many phenomena occurring in laser arrays, it cannot easily be adapted to include the effects of optical polarisation or electron spin that are often relevant in vertical cavity lasers. For this purpose, the spin flip model (SFM) \cite{ san1995light} is well-established as the method of choice, and has been successfully extended to model mutually coupled VCSELs by adding delayed optical injection terms \cite{vicente2006bistable}. This approach has been successfully applied recently to proposed applications of mutually coupled VCSELs in secure key distribution based on chaos synchronization \cite{jiang2017secure} and reservoir computing based on polarization dynamics \cite{guo2019four}.  

Spin-VCSELs, where the polarisation and dynamics can be controlled by the injection of spin-polarised carriers, have recently attracted considerable attention since very high-speed ($>$200 GHz) modulation has been demonstrated \cite{lindemann2019ultrafast} by applying mechanical stress to increase the birefringence. In the present contribution we explore some of the dynamics predicted for coupled pairs of spin-VCSELs based on a newly-developed theoretical treatment \cite{ vaughan2019analysis} that extends the SFM to apply to VCSEL arrays. This approach, which uses normal modes rather than CMT, accounts accurately for instantaneous coupling via evanescent fields or leaky waves. It is therefore able to model the details of the optical guidance in the spin-VCSELs and effects of varying the spacing between them, thus going beyond the description offered by adding optical injection terms to the conventional SFM. The next section gives a brief summary of this treatment leading to a set of rate equations. Subsequent sections deal with results, discussion and conclusions.

\section{Double-guided structure}\label{sec:double}

In Ref.~\cite{vaughan2019analysis}, a general set of rate equations for any number of coupled lasers with an arbitrary waveguide geometry and any number of optical modes, including the polarisation was derived from Maxwell's equations and the optical Bloch equations. In this model, the geometry of the waveguides is introduced through the introduction of \emph{overlap factors}, defined by

\begin{equation}
 \Gamma_{kk'}^{(i)} \equiv \int_{(i)} \Phi_{k}(\mathbf{r})\Phi_{k'}(\mathbf{r})~d\mathbf{r} \label{eq:Gamma_kk}
\end{equation}

\noindent where $k$ and $k'$ label the modes, $\Phi_{k}(\mathbf{r})$ is the spatial profile of the $k$th mode and the integral is over the $(i)$th guide. In fact, \eqref{eq:Gamma_kk} represents a simplified model for which the gain is assumed to be uniform over a guide and zero elsewhere. The mathematical model of Ref.~\cite{vaughan2019analysis} allows for a more general treatment, although this would greatly increase the complexity of the numerical solution. In an earlier work~\cite{vaughan2019stability}, we showed that the dynamics of coupled lasers in slab guides could be very sensitive to these overlap factors and stressed their importance.   

\begin{figure}[!ht]
\centering
\includegraphics[width=0.8\textwidth]{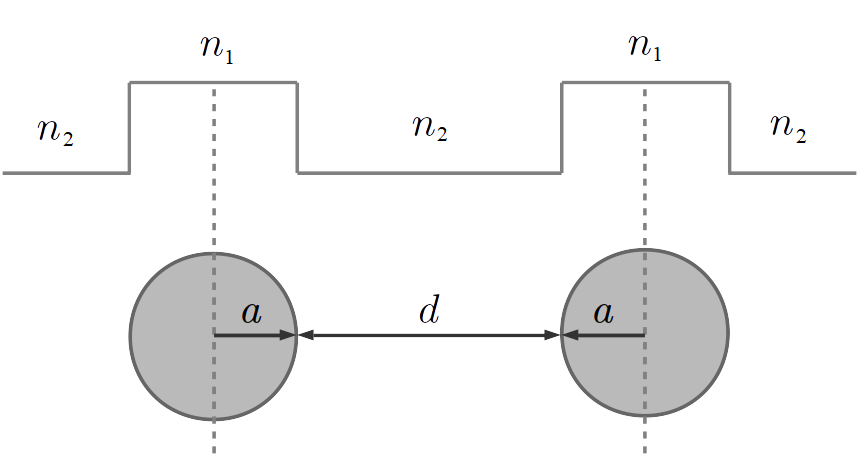}
\caption{\label{fig:waveguides} Circular guides of radius $a$ and edge-to-edge separation $d$. In this work, we set $a = 4~\mu\mathrm{m}$ and allow $d$ to be variable.}
\end{figure}

In the present work, we consider the particular case of double-guided structures consisting of two identical circular guides of radius $a = 4~\mu\mathrm{m}$, as illustrated in Fig.~\ref{fig:waveguides}. Note that in this paper, we take the edge-to-edge separation to be $d$ (rather than 2$d$ as in Ref.~\cite{vaughan2019analysis}). We choose values of the cladding refractive index $n_{2}$ and the refractive index in the guides $n_{1}$ such that, for the operating wavelength of $\lambda = 1.3~\mu\mathrm{m}$, there are only two supported modes with even (for the lower order mode) and odd parity. We shall refer to these as the \emph{symmetric} and \emph{anti-symmetric} modes and denote them by $k = s, a$ respectively. The values we choose are $n_{2} = 3.4$ and $n_{1} = 3.400971$. By restricting the number of solutions in this way with such small differences in refractive index, the evanescent tails of the optical modes tend to extend into the cladding regions to a significant extent. 

\begin{figure}[!ht]
\centering
\includegraphics[width=0.8\textwidth]{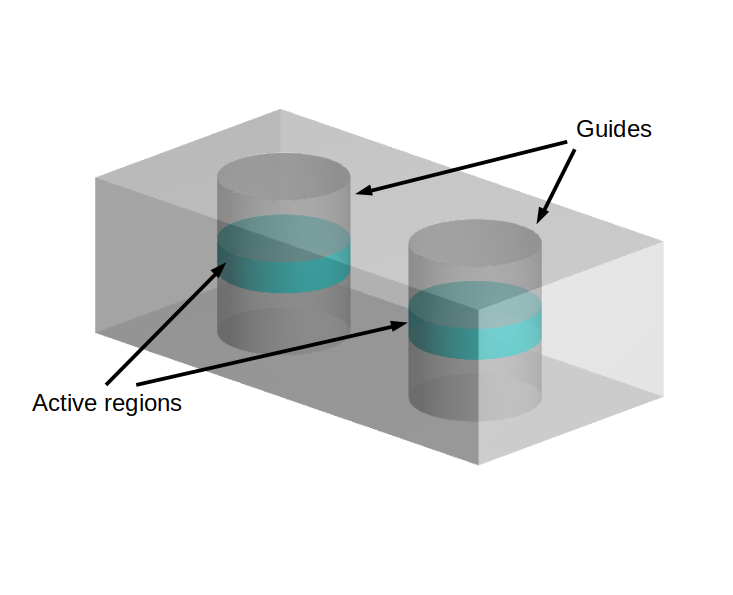}
\caption{\label{fig:VCSELS} A 3D schematic of two coupled circular waveguides encapsulating the essence of the application to a pair of VCSEL cavities. Shown are the cylindrical waveguide regions incorporating the active areas. Pumping is assumed to be confined to these regions. Note that we have omitted the Bragg stack mirrors and substrate from this figure.}
\end{figure}

An slightly more intuitive sketch of the guiding arrangement, illustrating how such a guiding configuration relates to VCSELs is shown in Fig~\ref{fig:VCSELS}. This extends the view of Fig.~\ref{fig:waveguides} into the propagation direction of the light and indicates, schematically, the active regions. To keep the diagram simple, no attempt has been made to add in further detail such as the Bragg stack mirrors. Due to their equivalence at this level of abstraction, we shall use the terms `guide' and `laser' interchangeably throughout the text. 

Limiting the number of modes to two, the notation required to denote the overlap factors may be specialised. Denoting the guide by the superscript $(i)$, $\Gamma_{ss}^{(i)}$ is the overlap of the symmetric modes, $\Gamma_{aa}^{(i)}$ the overlap of the antisymmetric modes and $\Gamma_{sa}^{(i)}$ is the cross product. Note that, due to the symmetry of the guides, we always have $\Gamma_{sa}^{(1)} = -\Gamma_{sa}^{(2)}$, due to the parity of the modes. Moreover, as the separation $d$ between them increases, we have $\Gamma_{sa}^{(i)} \to 0$ and $\Gamma_{ss}^{(i)} \to \Gamma_{aa}^{(i)} \to \Gamma_{S}/2$, where $\Gamma_{S}$ is the optical confinement factor of an isolated guide. The factor of 1/2 arises since the modes are normalised over all space, which includes 2 guides.

\subsubsection{Normalised rate equations}
The general form of the normal mode model and its reduction to the double-guided structure in dimensional and normalised form are derived in Ref~\cite{vaughan2019analysis}. Here we shall just quote the normalised form used in our numerical calculations. The model has 11 independent variables: the spin-polarised carrier concentrations in each guide $M_{\pm}^{(i)}$, where $i \in \{1,2\}$ labels the guide and $+/-$ labels the spin up / down components respectively; the optical amplitudes in each guide $A_{i,\pm}$, where $+/-$ labels the right-circularly / left-circularly polarised components respectively and three phase variables $\phi_{21++}$, $\phi_{21--}$ and $\phi_{11+-}$. The $\phi_{21\pm\pm}$ are the phase differences between $A_{2,\pm}$ and $A_{1,\pm}$, which we shall refer to as the \emph{spatial phase}. This is the phase of the coupled mode model of Ref~\cite{adams2017effects}. The variable $\phi_{11+-}$ is the phase difference between $A_{1,+}$ and $A_{1,-}$, which is the phase referred to in the literature of the SFM. We shall call this the \emph{polarisation phase}. A fourth phase variable $\phi_{11+-}$ is related to the other three via $\phi_{22+-} = \phi_{21++} - \phi_{21--} + \phi_{11+-}$.

Note that the $A_{i,\pm}$ are \emph{not} the amplitudes of the modal solutions of the Helmholtz equation but rather `composite modes' defined in terms of a superposition of the actual modal solutions (symmetric and anti-symmetric) to better exploit the symmetry of the waveguide. Specifically, these become the amplitudes of the local solutions in isolated guides as the separation between them is increased to infinity, retaining close similarity at nearer distances. Hence, they offer a more intuitive, physical representation of the optical field in each guide. The actual normal modes may be reconstructed from the composite modes and the phases using the procedure described in the appendix of Ref~\cite{vaughan2019analysis}.

For convenience of formulation in the double-guided structure model, we introduce new $\Gamma$ terms defined in terms of the optical overlap factors by

\begin{equation}
 \Gamma_{\pm}^{(i)} = \frac{\Gamma_{ss}^{(i)} + \Gamma_{aa}^{(i)} \pm 2\Gamma_{sa}^{(i)}}{2} \label{eq:Gamma_pm}
\end{equation}

\noindent and

\begin{equation}
 \Delta\Gamma^{(i)} = \frac{\Gamma_{ss}^{(i)} - \Gamma_{aa}^{(i)}}{2}. \label{eq:DGamma}
\end{equation}

\noindent Using these, we introduce further new variables defined via

\begin{equation}
 M_{12\pm} = \frac{\Gamma_{+}^{(1)}M_{\pm}^{(1)} + \Gamma_{+}^{(2)}M_{\pm}^{(2)}}{\Gamma_{S}}, \label{eq:M12}
\end{equation}

\begin{equation}
 M_{21\pm} = \frac{\Gamma_{-}^{(1)}M_{\pm}^{(1)} + \Gamma_{-}^{(2)}M_{\pm}^{(2)}}{\Gamma_{S}} \label{eq:M21}
\end{equation}

\noindent and 

\begin{equation}
 \Delta M_{\pm}  = \frac{\Delta\Gamma^{(1)}M_{\pm}^{(1)} + \Delta\Gamma^{(2)}M_{\pm}^{(2)}}{\Gamma_{S}}, \label{eq:DM}
\end{equation}

\noindent in terms of which the optical rate equations are more concisely written.

The normalised carrier rate equations are

\begin{equation}
\frac{\partial M_{\pm}^{(i)}}{\partial t} = \gamma\left[\eta_{\pm}^{(i)} - \left(1 + \mathcal{I}_{\pm}^{(i)}\right)M_{\pm}^{(i)}\right] - \gamma_{J}\left(M_{\pm}^{(i)} - M_{\mp}^{(i)}\right), \label{eq:dNdt_norm}
\end{equation}

\noindent where $\eta_{\pm}^{(i)}$ are the polarised pumping rates in each guide, $\gamma = 1/\tau_{N}$ is the inverse of the carrier lifetime $\tau_{N}$, $\gamma_{J}$ is the spin relaxation rate and the polarised components of the optical intensity in each guide are given by

\begin{align}
\mathcal{I}_{\pm}^{(i)} &= \frac{\Gamma_{+}^{(i)}}{\Gamma_{S}}|A_{1,\pm}|^{2} + 2\frac{\Delta\Gamma^{(i)}}{\Gamma_{S}}|A_{1,\pm}||A_{2,\pm}|\cos(\phi_{21\pm\pm}) \nonumber \\
&+ \frac{\Gamma_{-}^{(i)}}{\Gamma_{S}}|A_{2,\pm}|^{2}. \label{eq:Inorm}
\end{align}

\noindent Note that in the normalised form of the SFM, the effective spin relaxation rate $\gamma_{s} = \gamma + 2\gamma_{J}$ is often used.

The normalised optical rate equations are
 
\begin{align}
\frac{\partial |A_{1,\pm}|}{\partial t} &= \kappa\left(M_{12\pm} - 1\right)|A_{1,\pm}| + \left[\kappa\Delta M_{\pm}\left(\cos(\phi_{21\pm\pm}) - \alpha\sin(\phi_{21\pm\pm})\right) - \mu\sin(\phi_{21\pm\pm})\right]|A_{2,\pm}| \nonumber \\
&- \left[\gamma_{a}\cos(\phi_{11+-}) \pm \gamma_{p}\sin(\phi_{11+-})\right]|A_{1,\mp}|, \label{eq:dA1pmdt}
\end{align}

\begin{align}
\frac{\partial |A_{2,\pm}|}{\partial t} &= \kappa\left(M_{21\pm} - 1\right)|A_{2,\pm}| + \left[\kappa\Delta M_{\pm}\left(\cos(\phi_{21\pm\pm}) + \alpha\sin(\phi_{21\pm\pm})\right) + \mu\sin(\phi_{21\pm\pm})\right]|A_{1,\pm}| \nonumber \\
&- \left[\gamma_{a}\cos(\phi_{22+-}) \pm \gamma_{p}\sin(\phi_{22+-})\right]|A_{2,\mp}|, \label{eq:dA2pmdt}
\end{align}

\begin{align}
\frac{\partial\phi_{21\pm\pm}}{\partial t} &= \kappa\alpha\left(M_{21\pm} - M_{12\pm}\right) +\mu\cos(\phi_{21\pm\pm})\left(\frac{|A_{1,\pm}|}{|A_{2,\pm}|} -\frac{|A_{2,\pm}|}{|A_{1,\pm}|}\right) \nonumber \\
&+ \kappa\Delta M_{\pm}\left[\alpha\cos(\phi_{21\pm\pm})\left(\frac{|A_{1,\pm}|}{|A_{2,\pm}|} - \frac{|A_{2,\pm}|}{|A_{1,\pm}|}\right) - \sin(\phi_{21\pm\pm})\left(\frac{|A_{1,\pm}|}{|A_{2,\pm}|} + \frac{|A_{2,\pm}|}{|A_{1,\pm}|}\right)\right] \nonumber \\
&+ \gamma_{p}\left[\cos(\phi_{11+-})\frac{|A_{1,\mp}|}{|A_{1,\pm}|} - \cos(\phi_{22+-})\frac{|A_{2,\mp}|}{|A_{2,\pm}|}\right] \mp \gamma_{a}\left[\sin(\phi_{11+-})\frac{|A_{1,\mp}|}{|A_{1,\pm}|} - \sin(\phi_{22+-})\frac{|A_{2,\mp}|}{|A_{2,\pm}|}\right], \label{eq:dphi_21pp_norm}
\end{align}

\noindent and

\begin{align}
\frac{\partial\phi_{11+-}}{\partial t}  &= \kappa\alpha\left(M_{12+} - M_{12-}\right) + \mu\left(\cos(\phi_{21++})\frac{|A_{2,+}|}{|A_{1,+}|} - \cos(\phi_{21--})\frac{|A_{2,-}|}{|A_{1,-}|}\right) \nonumber \\
&+ \kappa\Delta M_{+}\left(\alpha\cos(\phi_{21++}) + \sin(\phi_{21++})\right)\frac{|A_{2,+}|}{|A_{1,+}|} - \kappa\Delta M_{-}\left(\alpha\cos(\phi_{21--}) +\sin(\phi_{21--})\right)\frac{|A_{2,-}|}{|A_{1,-}|} \nonumber \\
&+ \gamma_{a}\sin(\phi_{11+-})\left(\frac{|A_{1,+}|}{|A_{1,-}|} + \frac{|A_{1,-}|}{|A_{1,+}|}\right) + \gamma_{p}\cos(\phi_{11+-})\left(\frac{|A_{1,+}|}{|A_{1,-}|} - \frac{|A_{1,-}|}{|A_{1,+}|}\right) \label{eq:dphi_11pm_norm}
\end{align}

\noindent The parameters of the optical model are the linewidth enhancement factor $\alpha$, the cavity loss rate $\kappa$, the dichroism rate $\gamma_{a}$, the birefringence rate $\gamma_{p}$ and the coupling coefficient $\mu$. Note that $\mu$ is given in terms of the modal frequencies by~\cite{marom1984relation, vaughan2019analysis}

\begin{equation}
 \mu = \frac{\nu_{s} - \nu_{a}}{2}, \label{eq:mu}
\end{equation}

\noindent for the symmetric ($s$) and anti-symmetric ($a$) modes found from solution of the Helmholtz equation for the waveguiding structure.

It will be convenient to define the \emph{pump ellipticity} in the $i$th guide in terms of the right and left circular polarised pumping rates $\eta_{+}^{(i)}$ and $\eta_{-}^{(i)}$ by

\begin{equation}
 P^{(i)} = \frac{\eta_{+}^{(i)} - \eta_{-}^{(i)}}{\eta_{+}^{(i)} + \eta_{-}^{(i)}}. \label{eq:Pi}
\end{equation}

\noindent Similarly, we may define the output optical ellipticity in the $(i)$th guide via

\begin{equation}
\varepsilon^{(i)} = \frac{|A_{i,+}|^{2} - |A_{i,-}|^{2}}{|A_{i,+}|^{2} + |A_{i,-}|^{2}}. \label{eq:optellip_i} 
\end{equation}

\noindent We describe this as the `modal' ellipticity since it is in terms of the composite mode amplitudes. Although this is defined for each guide, there is a spatial dependence beyond this. The actual ellipticity we would measure is given in terms of the spatially dependent components of the optical intensity via 

\begin{equation}
 \varepsilon(x,y) = \frac{\mathcal{I}_{+}(x,y) - \mathcal{I}_{-}(x,y)}{\mathcal{I}_{+}(x,y) + \mathcal{I}_{-}(x,y)}, \label{eq:spatial_ellip}
\end{equation}

\noindent where the $ \mathcal{I}_{\pm}(x,y)$ are given in terms of the normal mode amplitudes $A_{k,\pm}$ by

\begin{equation}
 \mathcal{I}_{\pm}(x,y) = \left|A_{s,\pm}\Phi_{s}(x,y) + A_{a,\pm}\Phi_{a}(x,y)\right|^{2}. \label{eq:spatial_I}
\end{equation}

\noindent Here, we have used the subscript $k$ to distinguish these as the amplitudes of the \emph{modal} solutions of the waveguide (i.e. the solutions of the Helmholtz equation) rather than the `composite modes' used elsewhere in this paper (as discussed above), which are denoted by the subscript $i$. See Ref~\cite{vaughan2019analysis} for further details of this calculation.

\section{Results and discussion}\label{sec:results}
\subsection{Stability boundaries}

\begin{figure}[!ht]
\centering
\includegraphics[width=1.0\textwidth]{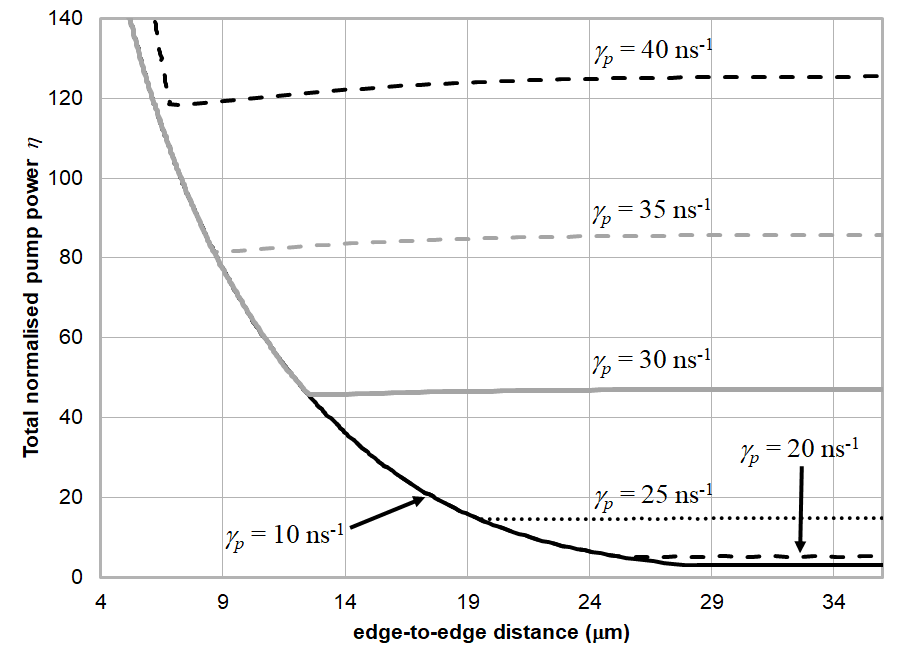}
\caption{\label{fig:inphaseBounds} Stability boundaries in the $\eta^{(i)} = \eta_{+}^{(i)} + \eta_{-}^{(i)}$ verses edge-to-edge distance $d$ plane for circular guides equally pumped with a pump ellipticity of $P^{(i)}= 0.0$ for different values of the birefringence rate $\gamma_{p}$.}
\end{figure}

The dynamics of pairs of laterally-coupled lasers with circular guides of radius $a = 4~\mu\mathrm{m}$ have been investigated by plotting stability boundaries in the $\eta^{(i)} - d$ plane, where $\eta^{(i)}= \eta_{+}^{(i)} + \eta_{-}^{(i)}$ is the total normalised pumping rate in either guide and $d$ is the edge-to-edge guide separation (Fig.~\ref{fig:inphaseBounds}). These plots are topologically equivalent to the scheme of $\Lambda/\Lambda_{th}$ v $d/a$ diagrams used in Refs~\cite{vaughan2019stability} and \cite{vaughan2019analysis}, where $\Lambda$ and $\Lambda_{th}$ are the total pump power and threshold pump respectively. Here, because we may vary the pump ellipticity in each guide independently, $\Lambda_{th}$ is not well defined and so represents an inaccurate measure. 

A similar stability map, in terms of $\Lambda/\Lambda_{th}$ v $d/a$ has been shown for the non-polarised case using the coupled mode model in Fig.~6 of Ref~\cite{adams2017effects}. A remaining discrepancy between the results of the coupled mode model and the present work is due to the sensitivity of the dynamics to the overlap factors. It was shown in Ref~\cite{vaughan2019stability} that, taking the asymptotic values of the overlap factors as the guide separation tended to infinity, the stability map for the non-polarised case reproduced that of the coupled-mode treatment in Ref~\cite{adams2017effects} exactly. This would then correspond to a birefringence rate of $\gamma_{p} = 0~\mathrm{ns}^{-1}$, which is almost indistinguishable from the case of $\gamma_{p} = 10~\mathrm{ns}^{-1}$ plotted in Fig.~\ref{fig:inphaseBounds}. 

For all the stability boundaries investigated here, we keep the total normalised pumping rate $\eta$ the same in each guide and so may be conveniently plotted in the  $\eta^{(i)}- d$ plane. In the regions of instability, we typically see oscillatory behaviour of the type reported in Section~\ref{sec:oscillations}. 

\begin{figure}[!ht]
\centering
\includegraphics[width=1.0\textwidth]{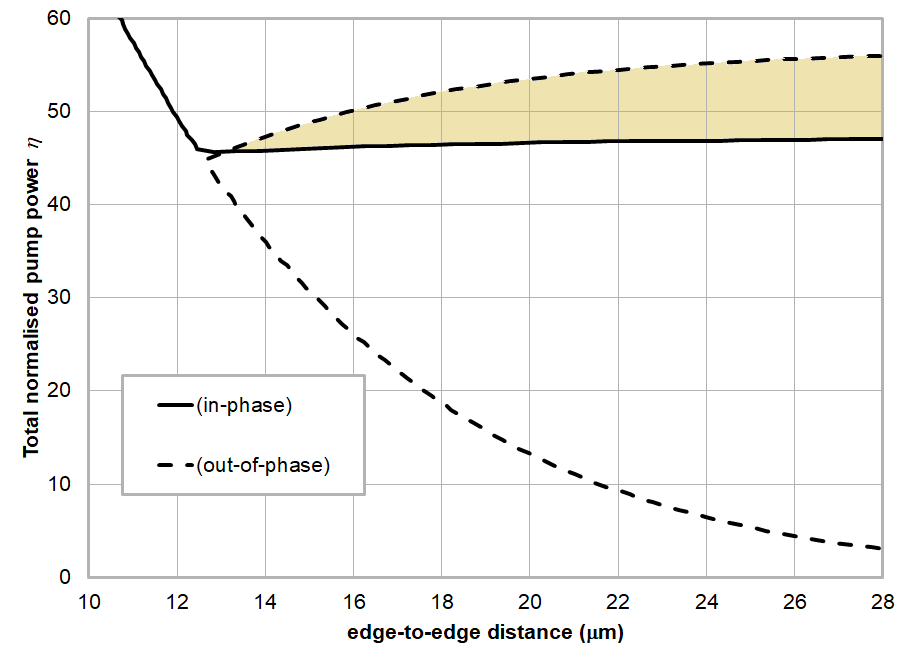}
\caption{\label{fig:bistableP00} Stability boundaries for $\gamma_{p} = 30~\mathrm{ns}^{-1}$ in the $\eta^{(i)}= \eta_{+} + \eta_{-}$ verses edge-to-edge distance $d$ plane for circular guides equally pumped with a pump ellipticity of $P^{(i)}= 0.0$. The solid lines are for the (polarisation) in-phase solutions and the dashed for the out-of-phase solutions. Stable in-phase solutions lie above the solid line, whilst stable out-of-phase solutions lie to the right and beneath the dashed line. The shaded area indicates the regions of bistability where both types of solution are stable.}
\end{figure}

\begin{table}
\caption{\label{tab:laser}Parameters used in numerical simulations. }

\begin{tabular}{llll}
\hline
\\
Parameter & Value & Unit & Description\\
\\
\hline\hline
\\
$\alpha$  & -2  &  & Linewidth enhancement\\
$\kappa$  & 70  & $\mathrm{ns}^{-1}$ & Cavity loss rate\\
$\gamma$  & 1  & $\mathrm{ns}^{-1}$ & Carrier loss rate\\
$\gamma_{a}$  & 0.1  & $\mathrm{ns}^{-1}$ & Dichroism rate\\
$\gamma_{s}$  & 100  & $\mathrm{ns}^{-1}$ & Effective spin relaxation rate\\
$N_{0}$  & $1.1\times10^{18}$  & $\mathrm{cm}^{-3}$ & Transparency density\\
$a_{diff}$  & $1.1\times10^{-15}$  & $\mathrm{cm}^{-1}$ & Differential gain\\
$n_{g}$  & 3.4  &  & Group refractive index\\
\\
\hline\hline
\\
\end{tabular}

Note, we tabulate the effective spin relaxation rate $\gamma_{s} = \gamma + 2\gamma_{J}$ to aid direct comparison with the SFM model. These are the same parameters as used in Ref.~\cite{vaughan2019analysis} except that in this paper we vary the birefringence rate $\gamma_{p}$.

\end{table}

In Ref.~\cite{vaughan2019analysis} stability boundaries were plotted for devices with a small birefringence rate $\gamma_{p}$ of $2~\mathrm{ns}^{-1}$, which gives very little coupling between the right and left circularly polarised components of the optical field. These gave rise to Hopf bifurcations qualitatively similar to the curve for  $\gamma_{p} = 10~\mathrm{ns}^{-1}$ shown in Fig.~\ref{fig:inphaseBounds}, up to around $d = 25~\mu\mathrm{m}$ (in these calculations, all other parameters have been kept the same as in Ref.~\cite{vaughan2019analysis} for the purposes of comparison). In this earlier work, the stable, steady state solutions found above the curve were termed `out-of-phase' solutions, in keeping with the terminology of the coupled mode model~\cite{adams2017effects}. In terms of the normal mode model, such out-of-phase solutions correspond to the anti-symmetric normal modes (at large separation, these tend to the solutions of isolated guides with a phase difference of $\pi$ between them, meaning the amplitudes are inverted). This phase relation is associated with the $\phi_{21\pm\pm}$ variables, i.e. at large separation $\phi_{21\pm\pm} = \pi$. Earlier, we designated this the  \emph{spatial phase} to distinguish it from the \emph{polarisation phase} associated with the $\phi_{ii+-}$ variables.

\begin{figure}[!ht]
\centering
\includegraphics[width=1.0\textwidth]{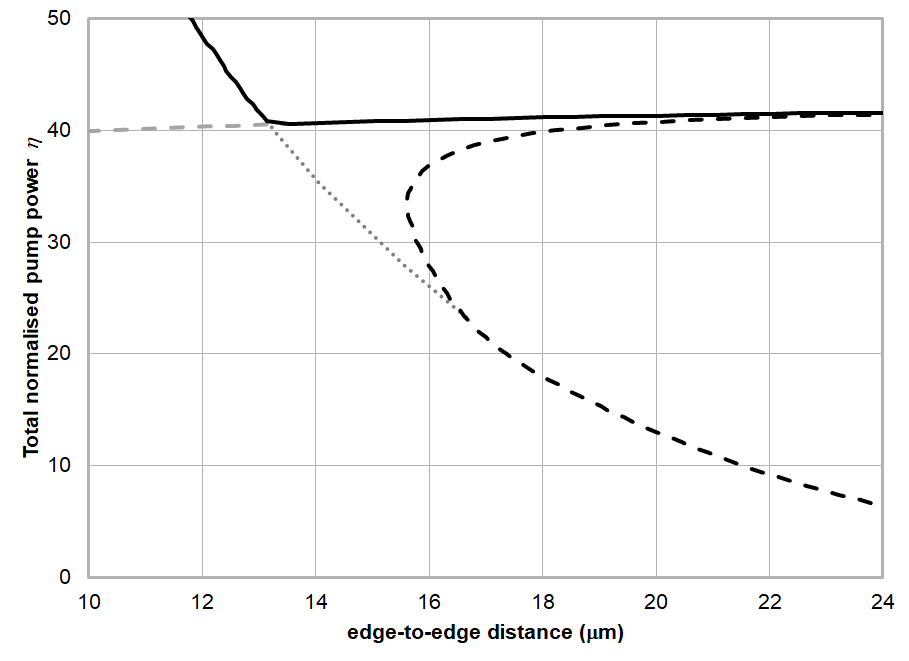}
\caption{\label{fig:bistableP02} Stability boundaries for $\gamma_{p} = 30~\mathrm{ns}^{-1}$ in the $\eta^{(i)}= \eta_{+} + \eta_{-}$ verses edge-to-edge distance $d$ plane for circular guides equally pumped with a pump ellipticity of $P^{(i)}= 0.2$. The grey lines show continuations of the Hopf bifurcation into the unstable region. Note that, unlike the other stability maps shown, in this case there is no region of bistability.}
\end{figure}

The graphs in Fig.~\ref{fig:inphaseBounds} are calculated for pump ellipticities of $P^{(i)}= P^{(1)} = P^{(2)} = 0$ (i.e. for linearly polarised pumps). In Ref~\cite{vaughan2019analysis} it was shown that, in general, the stability boundaries tended to move towards the origin as the pump ellipticity moves away from zero. In this work, we investigate the effect of increasing the birefringence rate $\gamma_{p}$, which has the effect of coupling power between the opposite circular components of the optical polarisation. Here, we see the emergence of a new stability boundary moving roughly horizontally across the plane and increasing in $\eta$ as $\gamma_{p}$ increases. These boundaries are plotted for the polarisation in-phase solutions, for which $\phi_{ii+-}$ is close to zero. These are characterised by the fact that the output optical ellipticity takes the same sign as the pump ellipticity. On the other hand, the ellipticity of the polarisation out-of-phase solutions, for which $\phi_{ii+-}$ is close to $\pi$, has the opposite sign to the pump ellipticity.  

\begin{figure}[!ht]
\centering
\includegraphics[width=1.0\textwidth]{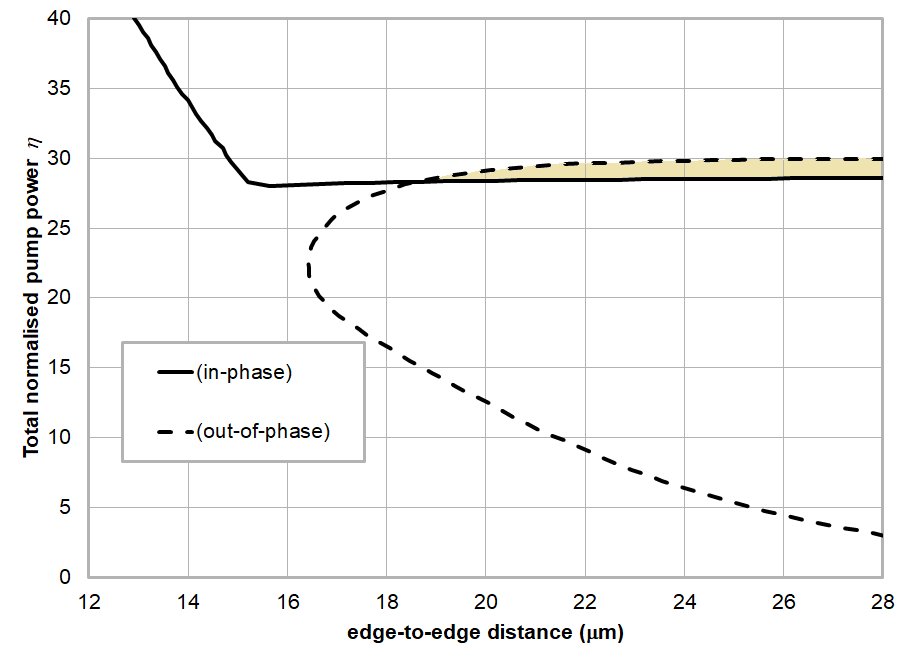}
\caption{\label{fig:bistableP04} Stability boundaries for $\gamma_{p} = 30~\mathrm{ns}^{-1}$ in the $\eta^{(i)}= \eta_{+} + \eta_{-}$ verses edge-to-edge distance $d$ plane for circular guides equally pumped with a pump ellipticity of $P^{(i)}= 0.4$ (details as for Fig.~\ref{fig:bistableP00}).}
\end{figure}

\begin{figure}[!ht]
\centering
\includegraphics[width=1.0\textwidth]{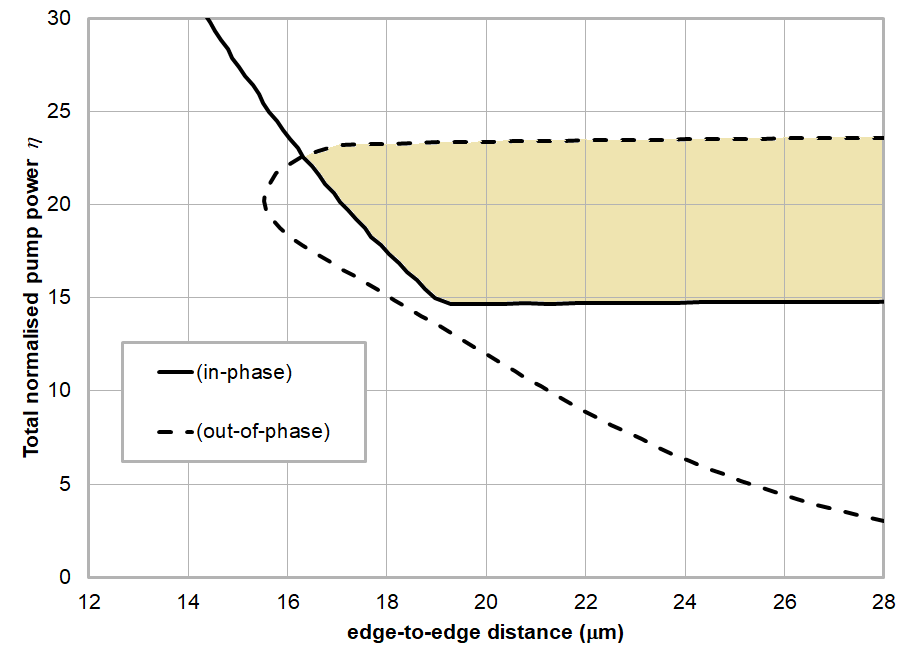}
\caption{\label{fig:bistableP06} Stability boundaries for $\gamma_{p} = 30~\mathrm{ns}^{-1}$ in the $\eta^{(i)}= \eta_{+} + \eta_{-}$ verses edge-to-edge distance $d$ plane for circular guides equally pumped with a pump ellipticity of $P^{(i)}= 0.6$ (details as for Fig.~\ref{fig:bistableP00}).}
\end{figure}

\begin{figure}[!ht]
\centering
\includegraphics[width=1.0\textwidth]{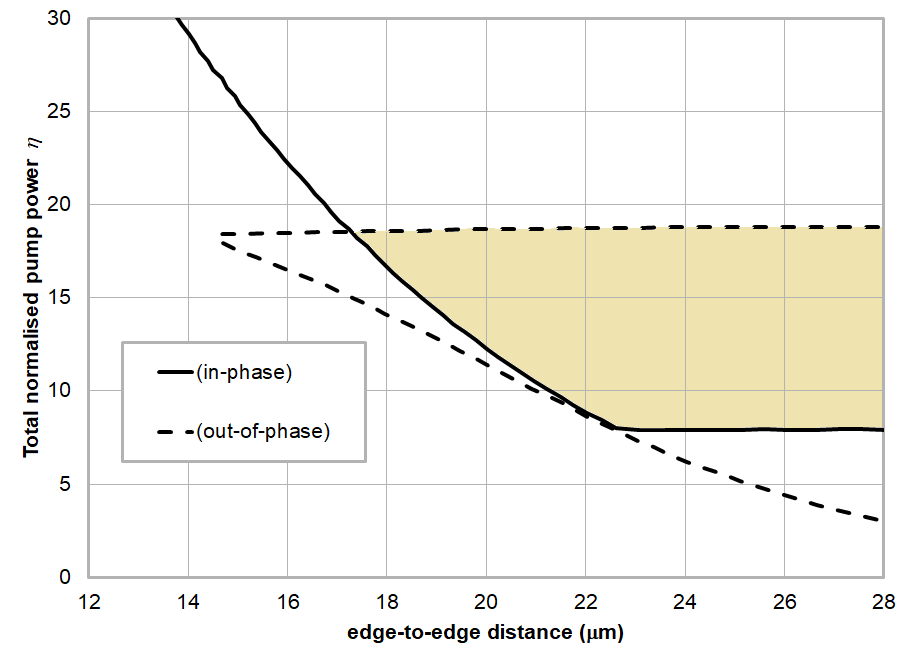}
\caption{\label{fig:bistableP08} Stability boundaries for $\gamma_{p} = 30~\mathrm{ns}^{-1}$ in the $\eta^{(i)}= \eta_{+} + \eta_{-}$ verses edge-to-edge distance $d$ plane for circular guides equally pumped with a pump ellipticity of $P^{(i)}= 0.8$ (details as for Fig.~\ref{fig:bistableP00}).}
\end{figure}

Stability boundaries for both in-phase and out-of-phase solutions for for $\gamma_{p} = 30~\mathrm{ns}^{-1}$ and values of  $P^{(i)}= P^{(1)} = P^{(2)}$ from 0 to 0.8 are shown in Figs~\ref{fig:bistableP00} to \ref{fig:bistableP08} (from here on, we shall be referring to the polarisation phase whenever we speak of in-phase or out-of-phase solutions without specific qualification). These show the out-of-phase stability boundaries as dashed lines with the stable solutions to the right of the curved borders and beneath the horizontal borders. Investigating the sharp kinks in the borders, we find that this is due to the continuation of Hopf bifurcations into the unstable regions. An example is shown in Fig.~\ref{fig:bistableP02} in the case of $P^{(i)}= 0.2$

A clear feature of these stability boundaries is that, in most cases, the out-of-phase boundary crosses that of the in-phase boundary creating regions of bistability where both types of solution are stable. These are shown as the shaded areas and suggest the possibility of optical switching between these stable states. This is investigated in the next sub-section, where we find that optical switching via pump power and / or ellipticity is indeed achievable.

In the case of $P^{(i)}= 0.2$, we note that, unlike the other cases shown, there is no region of bistability in domain plotted. At this point, however, we can offer no definitive explanation for this behaviour.

\subsection{Bistability}
\subsubsection{Switching both lasers together on pump power}

\begin{figure}[!ht]
\centering
\includegraphics[width=1.0\textwidth]{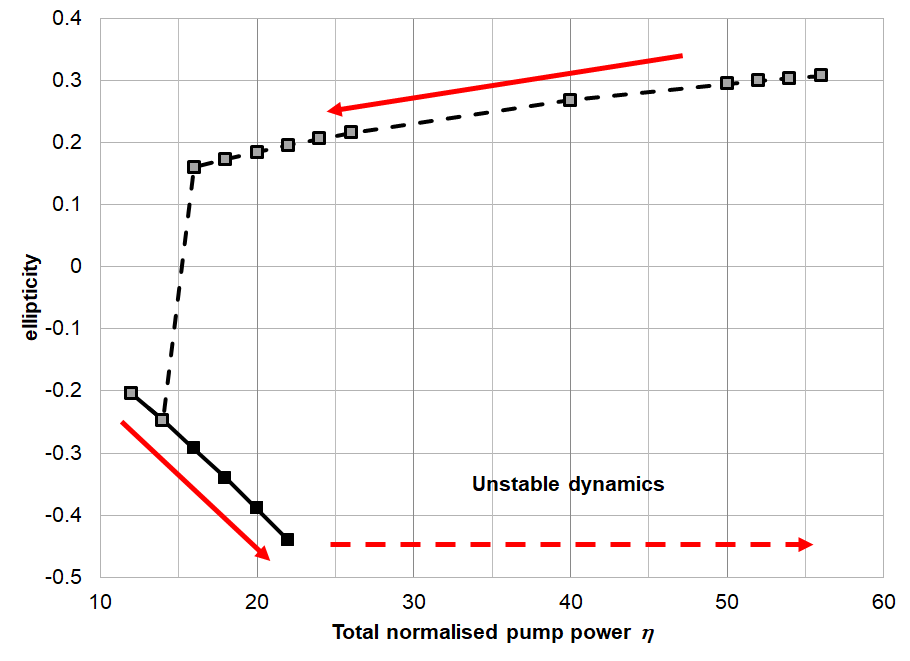}
\caption{\label{fig:bistability_P06_d20} Hysteresis curve of the ellipticity for equally pumped guides with  $\gamma_{p} = 30~\mathrm{ns}^{-1}$, an edge-to-edge separation of $d = 20~\mu\mathrm{m}$ and pump ellipticity  $P^{(i)}= 0.6$. The points trace out the dynamics as the total pump power $\eta$ is changed gradually in the direction of the arrows.}
\end{figure}

\begin{figure}[!ht]
\centering
\includegraphics[width=1.0\textwidth]{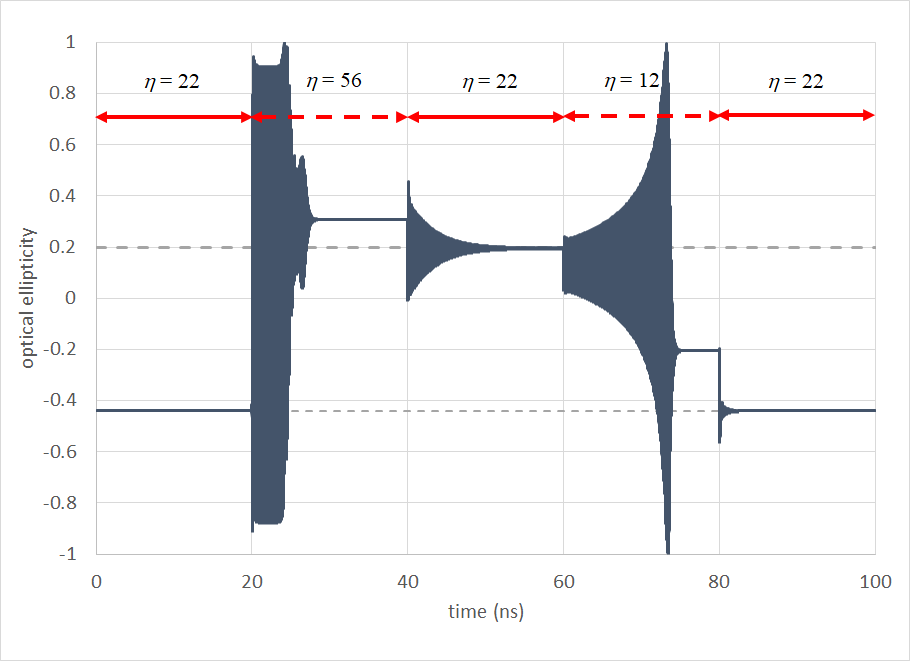}
\caption{\label{fig:time_steps} Time series showing the response of equally pumped guides with an edge-to-edge separation of $d = 20~\mu\mathrm{m}$ and pump ellipticity  $P^{(i)}= 0.6$, to stepped total pump powers of $\eta^{(i)}= 22, 56, 22, 12$ and 22 again. This demonstrates a mechanism of switching between two stable solutions for $\eta^{(i)}= 22$ with optical ellipticities of the opposite sign (indicated by the dashed lines).}
\end{figure}

\begin{figure}[!ht]
\centering
\includegraphics[width=1.0\textwidth]{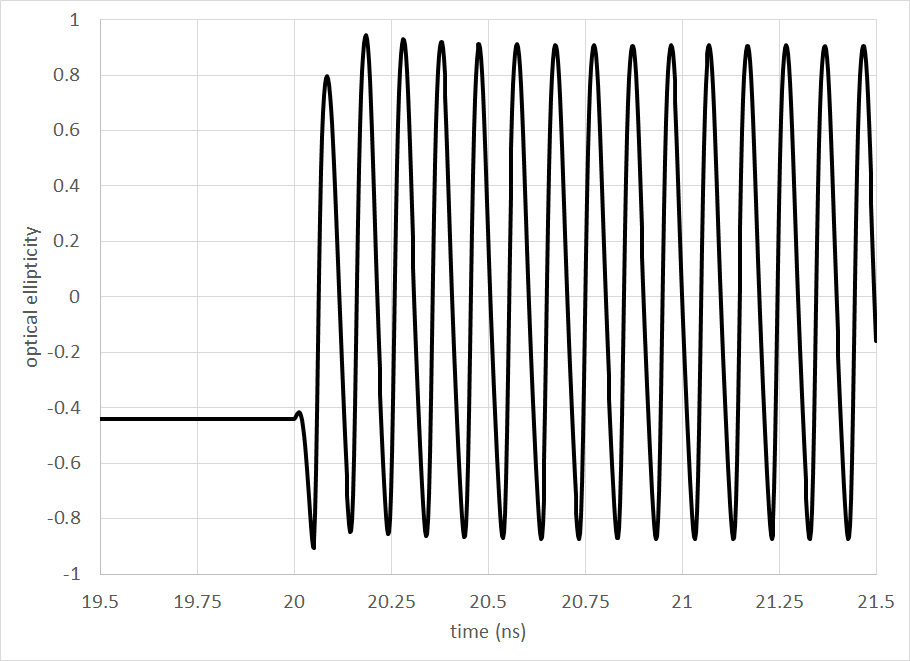}
\caption{\label{fig:time_steps_22_56} Detail of Fig.~\ref{fig:time_steps} between $t = 19.5~\mathrm{ns}$ and $t = 21.5~\mathrm{ns}$. The system starts in the out-of-phase steady-state solution for $\eta^{(i)}= 22$. As the pump power in each guide is stepped up $\eta^{(i)}= 56$, the ellipticity initially undergoes oscillations with an angular frequency of approximately 64~rad$\cdot$ns$^{-1}$ before settling down to a stable steady state in-phase solution.}
\end{figure}

\begin{figure}[!ht]
\centering
\includegraphics[width=1.0\textwidth]{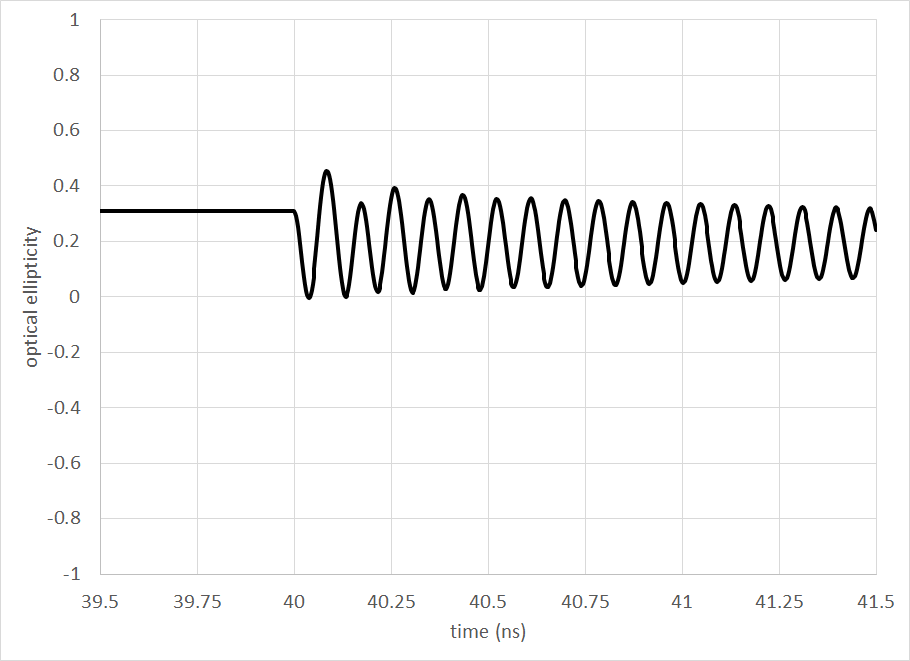}
\caption{\label{fig:time_steps_56_22} Detail of Fig.~\ref{fig:time_steps} between $t = 39.5~\mathrm{ns}$ and $t = 41.5~\mathrm{ns}$. Starting from the in-phase solution for $\eta^{(i)}= 56$, the pump power in each guide is stepped down to $\eta^{(i)}= 22$. The ellipticity initially undergoes oscillations with an angular frequency of approximately 72~rad$\cdot$ns$^{-1}$ and a much smaller amplitude than in Fig.~\ref{fig:time_steps_22_56} before decaying to the constant in-phase solution for $\eta^{(i)}= 22$.}
\end{figure}

\begin{figure}[!ht]
\centering
\includegraphics[width=1.0\textwidth]{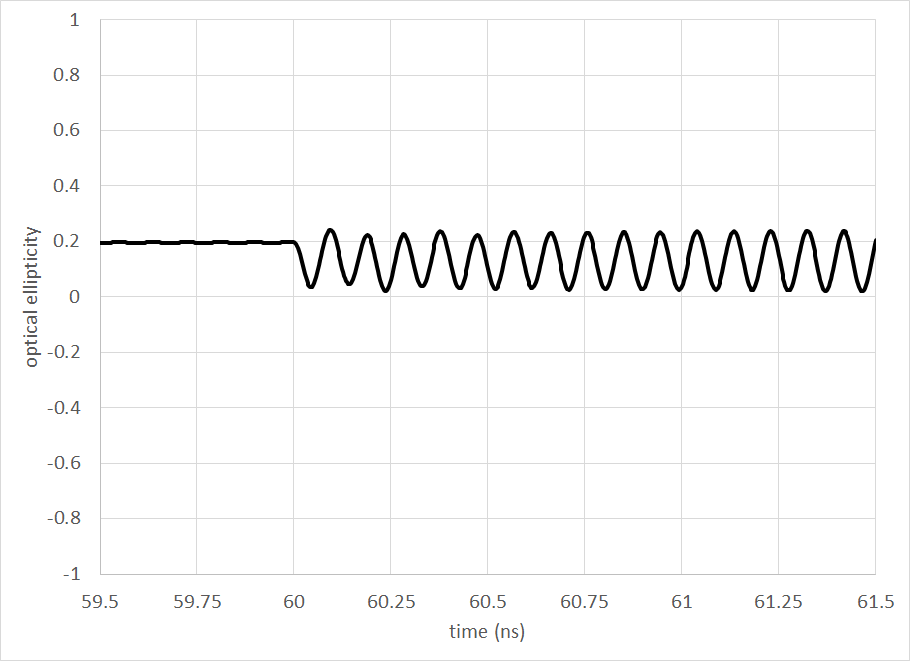}
\caption{\label{fig:time_steps_22_12} Detail of Fig.~\ref{fig:time_steps} between $t = 59.5~\mathrm{ns}$ and $t = 61.5~\mathrm{ns}$. From the in-phase solution with $\eta^{(i)}= 22$, the pump power is stepped down to $\eta^{(i)}= 12$, leading to oscillations in the ellipticity with an angular frequency of approximately 66~rad$\cdot$ns$^{-1}$. In this case the amplitude of the oscillation increases until it is varying between -1 and 1 before suddenly collapsing to a steady state out-of-phase solution.}
\end{figure}

\begin{figure}[!ht]
\centering
\includegraphics[width=1.0\textwidth]{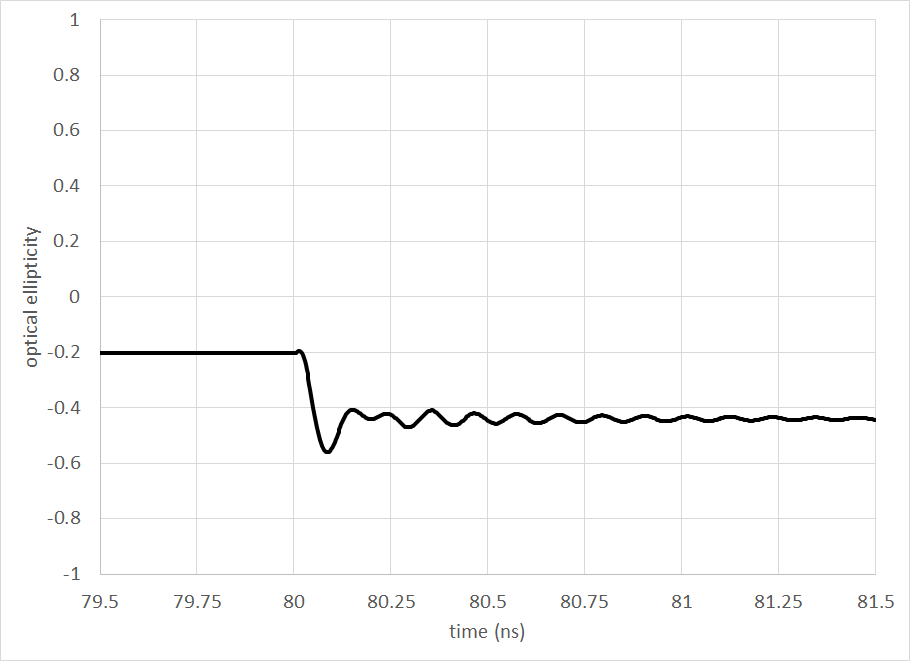}
\caption{\label{fig:time_steps_12_22} Detail of Fig.~\ref{fig:time_steps} between $t = 79.5~\mathrm{ns}$ and $t = 81.5~\mathrm{ns}$. With the system in an out-of-phase steady state solution at $\eta^{(i)}= 12$, the pump power is stepped up to $\eta^{(i)}= 22$. The ellipticity oscillates with an angular frequency of approximately 58~rad$\cdot$ns$^{-1}$ and then very rapidly decays to the out-of-phase steady state solution.}
\end{figure}

We have examined the dynamics within the bistable regions via time series solutions of the rate equations using the Runge Kutta method (technical details are given in Ref.~\cite{vaughan2019analysis}). Each time series is run for a simulation time of 400~ns for a given pump power and ellipticity in each guide. The output solutions at the end of each solution are then used as the intial conditions for the next simulation with different pumping parameters. In this way, we can see how the system behaves as we vary these parameters smoothly or in sharp jumps.

For an initial set of simulations, we keep the birefringence at $\gamma_{p} = 30~\mathrm{ns}^{-1}$, take the edge-to-edge separation to be $d = 20~\mu\mathrm{m}$ and the pump ellipticity in either guide to be $P^{(i)}= P^{(1)} = P^{(2)} = 0.6$. The stability boundaries in this case are shown in Fig.~\ref{fig:bistableP06}. We start the simulation with equal pump power $\eta^{(i)}= \eta^{(1)} = \eta^{(2)} = \eta_{+}^{(i)} + \eta_{-}^{(i)} = 12$. From Fig.~\ref{fig:bistableP06} we can see that this is in a region of instability for the in-phase solution but just on the edge of the stable region  for the out-of-phase solution. We then start increasing the pump power in both guides and track the modal output optical ellipticity $\varepsilon^{(i)}$, given by \eqref{eq:optellip_i}. This is shown in Fig.~\ref{fig:bistability_P06_d20}, where at $\eta^{(i)}= 12$ we have $\varepsilon^{(i)} = -0.2$ and track down to $\eta^{(i)}= 22$, $\varepsilon^{(i)} = -0.44$ following the direction of the red solid arrow. After this point, we enter into a region of unstable dynamics where the system fails to settle down to the in-phase steady state solution until the power reaches $\eta^{(i)}= 56$. This is indicated by the dashed red arrow. At this point, we track back, ramping down the power. This time, the system remains in the in-phase steady state solution all the way through the bistable region until it cross the Hopf bifurcation delimiting the in-phase dynamics and the system drops to the out-of-phase solution.

It is natural to ask whether we may obtain switching behaviour by applying step changes to the pump. To investigate this, we start the system off in an out-of-phase steady state solution with $\eta^{(i)}= 22$ in both lasers. This gives an output ellipticity of $\varepsilon^{(i)} = -0.44$. We then step up the power to $\eta^{(i)}= 56$ for a period of 20~ns. This settles down to a steady-state in-phase solution with $\varepsilon^{(i)} = 0.3$ after about 9~ns as shown in Fig~\ref{fig:time_steps}. After this, the power is dropped back down to $\eta^{(i)}= 22$. However, the system now settles down in an in-phase steady-state with $\varepsilon^{(i)} = 0.2$. Again, it takes aroung 9 to 10~ns for the system to settle to the steady-state solution. Following this, the power is further dropped to $\eta^{(i)}= 22$ and the system switches to an out-of-phase solution with $\varepsilon^{(i)} = -0.2$. Finally, stepping the power back up to $\eta^{(i)}= 22$, we arrive back at the out-of-phase solution with  $\varepsilon^{(i)} = -0.44$. Hence, we can use the pump power for the purposes of optical switching, with an overal switching time of around 20~ns in this case (giving a possible switching rate of around 8~MHz).

The switching dynamics are explored in more detail in Figs.~\ref{fig:time_steps_22_56} to \ref{fig:time_steps_12_22} on the sub-nanosecond time-scale. Fig.~\ref{fig:time_steps_22_56} shows the dynamics as the system is switched from the out-of-phase solution at $\eta^{(i)}= 22$ to the in-phase at $\eta^{(i)}= 56$. We see on this scale that the behaviour is oscillatory, varying between around $\varepsilon^{(i)} = -0.9$ to $\varepsilon^{(i)} =0.9$ with an angular frequency of approximately 64~ rad$\cdot$ns$^{-1}$ ($\sim$10~GHz). Figs.~\ref{fig:time_steps_56_22} to \ref{fig:time_steps_22_12} show the steps from  $\eta^{(i)}= 56$ to $\eta^{(i)}= 22$ and $\eta^{(i)}= 22$ to $\eta^{(i)}= 12$ respectively on the same scale, with similar angular frequencies of 72~rad$\cdot$ns$^{-1}$ ($\sim$11~GHz) and 66~rad$\cdot$ns$^{-1}$ ($\sim$11 GHz). In the final step from $\eta^{(i)}= 12$ to  $\eta^{(i)}= 22$ shown in Fig.~\ref{fig:time_steps_12_22}, the system settles down much faster. The angular frequency of the oscillations in this case is around 58~rad$\cdot$ns$^{-1}$ ($\sim$9.2~GHz). 

It may seem natural to seek an explanation for this oscillatory behaviour in terms of relaxation oscillations. We can explore this possibility using the expression for the angular frequency $\omega_{R}$ of damped oscillations given in Ref.~\cite{adams2017effects} derived from a stability analysis of the coupled mode model

\begin{equation}
\omega_{R}^{2} = 2\gamma\kappa\left(\eta - 1\right) - \gamma_{D}^{2}, \label{eq:RO}
\end{equation}

\noindent where $\gamma_{D}$ is the damping rate given by

\begin{equation}
\gamma_{D} = -\frac{\gamma\eta}{2}. \label{eq:damping}
\end{equation}

\noindent However, for values of $\eta = 56, 22$ and $12$, using \eqref{eq:RO} we obtain values of $\omega_{R} = 83, 53$ and 39~ns$^{-1}$ respectively, showing a strong dependence on the pump power $\eta$.

Instead, we note that in the analysis of spin-polarised VCSELS \cite{gahl1999polarization, li2017stability, lindemann2019ultrafast}, it has been found that the frequency of birefringence-induced oscillations was mainly determined by the birefringence rate $\gamma_{p}$, given approximately by $\gamma_{p}/\pi$ for large $\gamma_{p}$ (in GHz if $\gamma_{p}$ is given in ns$^{-1}$). In our case, we have $\gamma_{p} =$~30~ns$^{-1}$, giving $\gamma_{p}/\pi =$~9.5~GHz, which is very close to the observed frequency in the numerical simulations.

\subsubsection{Switching one laser via the other}
Having verified that is possible to switch the ellipticity of the lasers in the bistable region by varying the pump powers in each simultaneously, we next investigate the possibility of switching one laser purely by varying the pump on the other, hence via the coupling between them. The following is a proof of concept and is not supposed to represent the optimal conditions for such functionality.

\begin{figure}[!ht]
\centering
\includegraphics[width=1.0\textwidth]{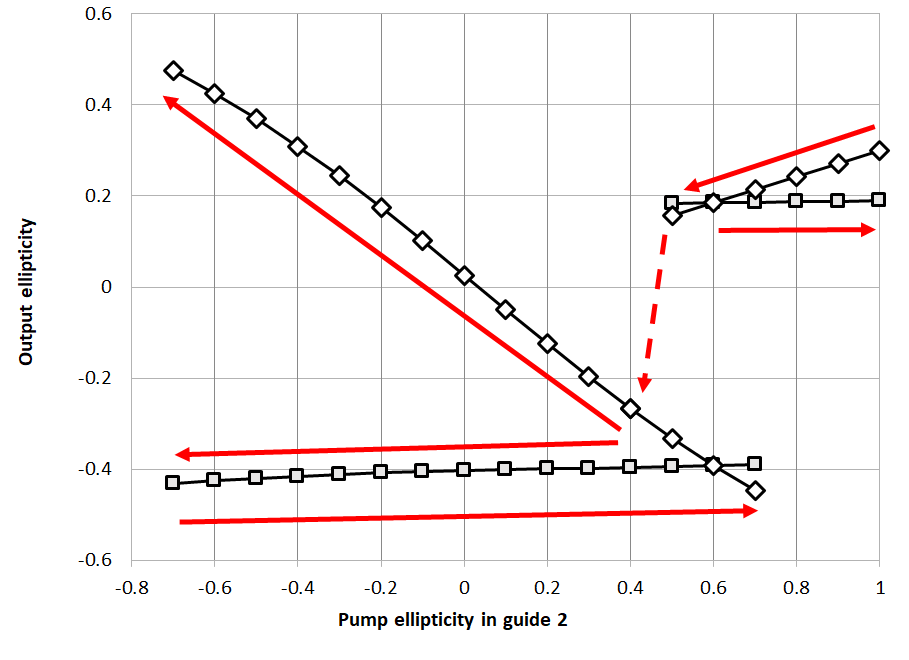}
\caption{\label{fig:P2_switch} Optical ellipticity switching by varying the pump ellipticity in guide 2. Here $\gamma_{p} = 30~\mathrm{ns}^{-1}$, $d = 18~\mu\mathrm{m}$,  $P^{(1)} = 0.6$ and $\eta^{(i)}= 20$. The grey squares show the optical ellipticity in guide 1 and the white diamonds show the ellipticity in guide 2, for which the pump ellipticity was directly varied. The red arrows show the sequence in which the pump ellipticity in guide 2,  $P^{(2)}$, was varied.}
\end{figure}

The edge-to-edge separation is taken to be a little shorter at $d = 18~\mu\mathrm{m}$ and for the initial investigation, the total pump power in either guide is held fixed at $\eta^{(i)}=20$. The birefringence is $\gamma_{p} = 30~\mathrm{ns}^{-1}$ as before. Initially, the pump ellipticity is set at $P^{(i)}= 0.6$ in both guides and the simulation is started with both lasers in the steady-state in-phase solution. $P^{(1)}$ is kept fixed throughout and $P^{(2)}$ is then varied, initially being increased to $P^{(2)} = 1$ and then reduced again to $P^{(2)} = 0.5$ (see Fig.~\ref{fig:P2_switch}). Throughout this range, both solutions remain in a stable in-phase solution. However, below $P^{(2)} = 0.5$, both lasers drop to an out-of-phase steady-state solution, which then varies smoothly as $P^{(2)}$ is reduced to -0.7. During this variation, the laser in guide (1) remains in an out-of-phase solution, whilst the ellipticity in guide (2) varies linearly from an out-of-phase solution to an in-phase solution. Beyond $P^{(2)} = -0.7$, neither laser settles to a steady-state.

As $P^{(2)} = -0.7$ is increased to $P^{(2)} = 0.7$, the ellipticity tracks back over its previous values and then continues to vary smoothly past the point where the in-phase solution dropped to the out-of-phase solution. These behaviours are shown in Fig.~\ref{fig:P2_switch} where the square points show the ellipticity in guide (1), the diamond points show the ellipticity in guide (2) and the red arrows indicate the directions in which $P^{(2)}$ is varied.

In fact, it is found that once the system is on the lower line of Fig.~\ref{fig:P2_switch} with guide (1) in an out-of-phase steady-state solution, it cannot be switched back to an in-phase state by varying $P^{(2)}$. This can only be achieved by varying the pump power. However, it can be achieved by only varying the pump power in laser (2), so the goal of switching one laser purely by coupling with the other is achievable. 

Specifically, we can use the following sequence: Starting with $\eta^{(i)}=20$ and  $P^{(i)} = 0.6$ in the in-phase solution, we have $\epsilon^{(1)} = 0.19$. Stepping $P^{(2)}$ to 0.4, $\epsilon^{(1)}$ drops to -0.40. Putting $P^{(2)}$ back to 0.6, $\epsilon^{(1)}$ changes very little, with $\epsilon^{(1)} = -0.39$. If we now step $\eta^{(2)}$ up to 60, the ellipticity in guide (1) then changes to $\epsilon^{(1)} = 0.19$. Dropping the power in guide (2) back down to 20, we end up again in the original in-phase solution with $\epsilon^{(1)} = 0.19$. For this particular set of parameters, the switching time is quite slow, taking around $100~\mathrm{ns}$ to settle down to the steady-state solutions.

\subsubsection{Oscillations in the ellipticity}\label{sec:oscillations}

\begin{figure}[!ht]
\centering
\includegraphics[width=1.0\textwidth]{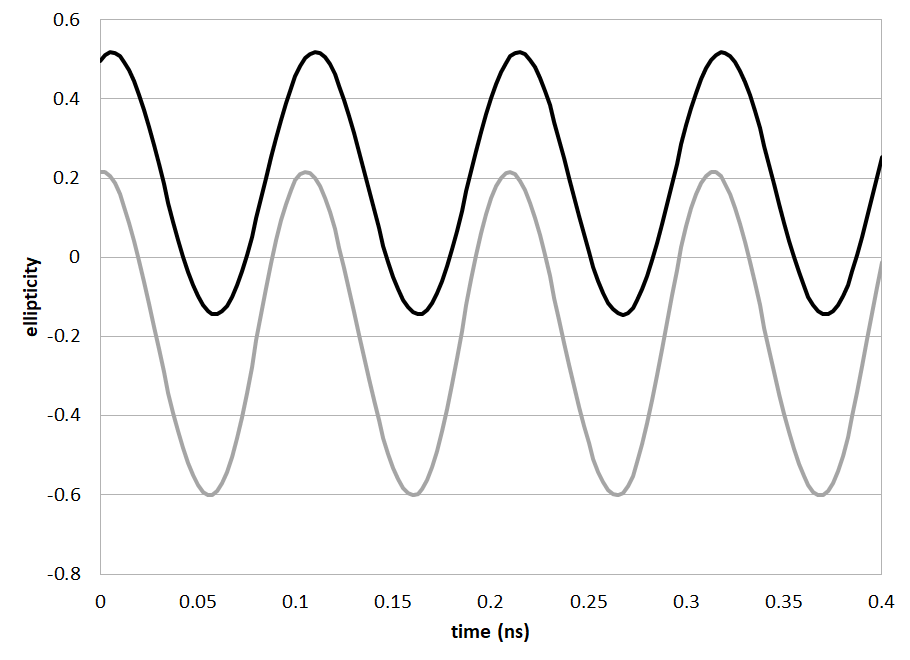}
\caption{\label{fig:oscillation_P203} Oscillations in the optical ellipticity for guide 1 (black) and guide 2 (grey) with a separation of $d = 20~\mu\mathrm{m}$. Here $\gamma_{p} = 30~\mathrm{ns}^{-1}$, $P^{(1)} = 0.6$, $P^{(2)} = 0.3$ and $\eta^{(i)}= 22$.}
\end{figure}

This switching behaviour on the basis of variation of $P^{(2)}$ does not occur under all conditions within a bistable region. At $d=20~\mu\mathrm{m}$, the variation in the output ellipticities is similar to that shown in Fig.~\ref{fig:P2_switch} except that there is drop from the steady-state in-phase solutions to the out-of-phase steady-state solutions as $P^{(2)}$ is reduced. Instead, the system becomes unstable with the ellipticity oscillating as shown in Fig.~\ref{fig:oscillation_P203} for $P^{(2)} = 0.3$.

\begin{figure}[!ht]
\centering
\includegraphics[width=1.0\textwidth]{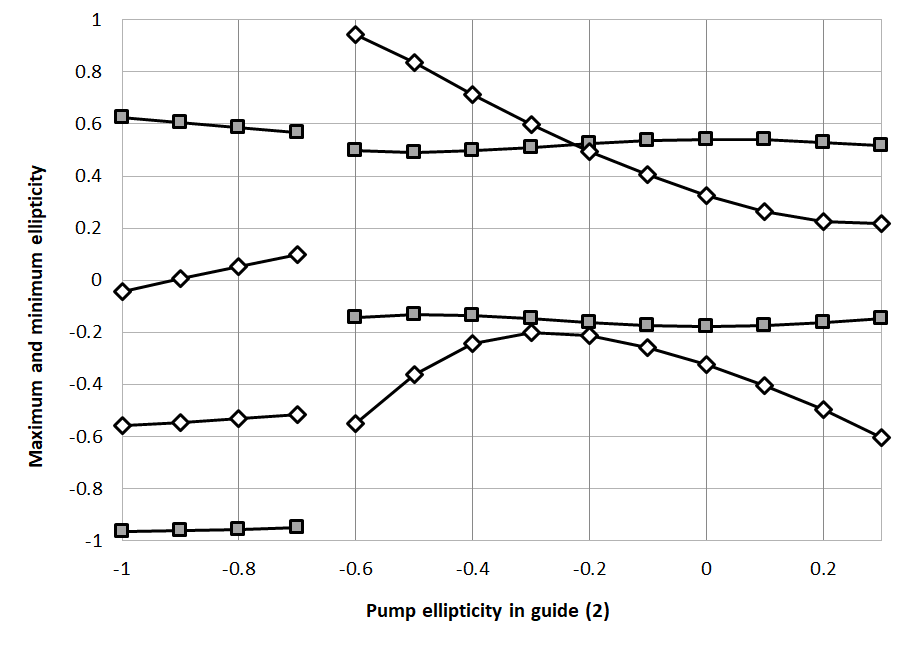}
\caption{\label{fig:min_max} Minima and maxima of the oscillations in the optical ellipticity for guide 1 (squares) and guide 2 (diamonds) with a separation of $d = 20~\mu\mathrm{m}$ as $P^{(2)}$ is varied. Here $\gamma_{p} = 30~\mathrm{ns}^{-1}$, $P^{(1)} = 0.6$,  and $\eta^{(i)}= 22$.}
\end{figure}

The time period for the oscillations shown in Fig.~\ref{fig:oscillation_P203} is approximately $T = 0.1~\mathrm{ns}$ (i.e. a frequency of 10~GHz). This does not change as $P^{(2)}$ is varied from 0.3 to -1, although the maxima and minima of the oscillations do. This variation is shown in Fig.~\ref{fig:min_max}. We note a qualitative break in behaviour between $P^{(2)} = -0.6$ and $P^{(2)} = -0.7$.

\section{Conclusions}
A recently-developed theory of evanescently-coupled pairs of spin-VCSELs has been applied to study the dynamics of structures with two identical circular cylindrical waveguides and realistic material parameters. Stability boundaries in the plane of total normalised pump power versus edge-to-edge spacing of the lasers have been presented for the cases of (1) zero pump polarization ellipticity with varying birefringence rate, and (2) fixed birefringence and varying pump ellipticity, with equal pump power in each laser for all cases. Boundaries for in-phase and out-of-phase solutions are found in terms of the spatial phase of the normal modes of the system. It is shown that intersection of these boundaries can give rise to sharp ‘kinks’ in the overall  stability boundaries for some pump ellipticities, whilst for others crossing of the in-phase and out-of-phase solutions can yield regions of bistability.
The dynamics of the coupled spin-VCSELs in the bistable regions have been examined by time series solutions of the rate equations. It is shown that it is possible to switch the output ellipticity of the lasers by varying the pump powers in each simultaneously. It is also possible to switch the output elllipticity of one laser by varying the pump ellipticity or pump power of the other, under certain operating conditions. For other conditions, however, values of the pump ellipticity of one laser can be found that produce oscillatory behaviour of the output ellipticities of both lasers. Thus, it has been demonstrated that evanescently-coupled pairs of spin-lasers can yield a rich variety of different dynamics. Further work is needed to explore the effects of varying material, device and operating parameters and hence to investigated potential applications of these dynamics.
\section*{Acknowledgement}
This research was funded by the Engineering and Physical Sciences Research Council (EPSRC) under grant No. EP/M024237/1.

\bibliography{spinVCSEL}

\end{document}